# Giant linearly-polarized photogalvanic effect and second harmonic generation in two-dimensional axion insulators

Ruixiang Fei, [1]  Wenshen Song, [1]  Li Yang [1,2,*]

[1] *Department of Physics, Washington University in St Louis, St Louis, Missouri 63130, United States*

[2] *Institute of Materials Science and Engineering, Washington University in St. Louis, St. Louis, Missouri 63130, United States*

[*]Email: lyang@physics.wustl.edu

## Abstract

The second-order nonlinear optical (NLO) processes, such as the photogalvanic effect and second-order harmonic generation (SHG), play crucial roles in probing and controlling light-matter interactions for energy and device applications. To date, most studies of second-order NLO processes focus on materials with broken spatial inversion symmetry, such as proper ferroelectrics and noncentrosymmetric Weyl semimetals. Nevertheless, inversion symmetry of Shubnikov groups can be broken via spin-ordering in centrosymmetric crystals. Unfortunately, these materials are less common, and their NLO responses are usually weak. Combining quantum perturbation theory and first-principles simulations, we predict a giant injection-current photogalvanic effect and SHG in a family of emerging axion insulators, the even septuple layers of $MnBi_2Te_4$ (MBT) materials that exhibit the zero-plateau quantum anomalous Hall (QAH) effect. Their amplitudes of injection current and SHG are about two orders of magnitude larger than those of widely used ferroelectrics, such as $BiFeO_3$ and $LiNbO_3$. Moreover, unlike the usual injection current observed under circularly-polarized light, the injection photocurrent of MBTs only emerges under linearly polarized light, making it convenient for device applications. These unique characters are from a combination effect of parity-time symmetry, three-fold rotation symmetry, and significant spin-orbit coupling. These enhanced NLO effects are valuable for characterizing subtle topological orders in QAH systems and also shed light on novel infrared photo-detector and photovoltaic applications based on magnetic topological materials.

*Introduction*: The past decade has witnessed explosive discoveries of the interplays between quantum phases and symmetries. The best-established examples include the electric polarization observed in solids with broken crystal inversion symmetry [1,2], chiral edge states in time-invariant topological insulators (TIs) [3,4], and Weyl fermions in time-reversal broken or inversion broken topological semimetals [5–7]. Meanwhile, light-matter interactions have been intensively applied for detecting material symmetries and corresponding quantum phases. Particularly, because nonlinear optical (NLO) processes are known for being tightly associated with symmetry breakings, remarkable second-order NLO responses have been extensively found in above-discussed materials. These include the high-voltage photovoltaic effect in ferroelectrics [8–12], enhanced shift-current bulk photovoltaics in two-dimensional (2D) ferroelectrics [13–15] and non-centrosymmetric TIs [16], and the quantized injection-current photogalvanic effect [17,18] and giant anisotropic second harmonic generation (SHG) [19] in Weyl semimetals. Importantly, all the above materials keep time-reversal symmetry and realize second-order NLO via breaking spatial inversion symmetry, leading to non-vanishing Berry phase and subsequent concept of electric polarization.

On the other hand, non-centrosymmetry in Shubnikov groups and possible second-order NLO responses can be realized through tuning spin degree of freedom [20–22] even with a zero Berry phase or curvature. Unfortunately, this type of NLO responses are usually weak and less common. The emerging magnetic topological materials may shed light on changing this wisdom. Particularly, the strong spin-orbit coupling (SOC), which is usually required in topological materials, may substantially break the $SU(2)$ spin-rotation symmetry and subsequent centrosymmetry, giving hope to enhanced second-order NLO processes.

Among numerous topological materials, MnBi$_2$Te$_4$ (MBT) may provide an ideal opportunity for realizing second-order NLO responses due to its significant SOC and rich topological properties under different symmetries and magnetic orderings. Bulk MBT is a large-gap antiferromagnetic (AFM) TI [23–28]. When thinning down to odd-number septuple layers, they exhibit an uncompensated AFM (uAFM) state. The quantum anomalous Hall effect (QAHE) was proposed [25–27,29,30] and confirmed in recent experiments [31–33]. Nevertheless, the even-number septuple layers of MBT is compensated AFM (cAFM) insulators with parity-time (*PT*) symmetry, leading to a zero-plateau QAHE (zQAHE) with a gapped chiral edge state [29,32]. As a result, the topological axion state emerges, showing a quantized topological magnetoelectric effect [34,35]. Because magnetic configurations can manifest various topological phases in MBT family, it will be highly appreciable if the magnetic and subsequent topological orders can be connected with second-order NLO responses.

In this letter, we predict enhanced photocurrent and SHG in the AFM zQAHE materials, i.e., the even-number septuple layers of MBT materials, including $MnBi_2Te_4$, $MnBi_2Se_4$, $MnSb_2Te_4$, $NiBi_2Te_4$, and $NiBi_2Se_4$. Because of broken magnetic inversion symmetry and enhanced SOC, the injection photocurrent and SHG of these materials are about one or two orders magnitude larger than those in well-known photovoltaic ferroelectrics and 2D non-centrosymmetric materials under linearly polarized light. With the help of *PT* and three-fold rotation symmetries, the injection-current photogalvanic effect of MBTs can only be observed upon linearly-polarized electromagnetic field, making it different for the widely observed circular photogalvanic effect in time-invariant systems.

*Structure and symmetries*: The tetradymite-type compounds of the MBT family crystallize in a rhombohedral structure with the space group $D_{3d}^5$(No. 166). Magnetic Mn atoms in each layer form triangular lattices with an ABC stacking along the out-of-plane direction. The ground state of bulk MBT is interlayer AFM with an intralayer FM ordering, forming the so-called A-type AFM ordering. Previous studies showed that bulk MBTs are AFM TIs [23,24,26,27,29]. Our first-principles simulation obtains the same ground-state structure and A-type AFM ordering. The details of the calculations are presented in the Supplemental Material [36–45].

There are several important symmetries of bulk MBTs. First, there is an inversion symmetry $P_1$ (centered at the point $P_1$ in Figure 1(a)). Second, the composition of spatial inversion and time-reversal ($P_2T$) symmetries is preserved. The inversion center is labeled by the point $P_2$ in Figure 1(a). Finally, non-symmorphic time-reversal symmetry ($S = T\tau$) is preserved as well, leading to a $Z_2$ classification [27,46], where τ is the half translation operator connecting the nearest spin-up and spin-down Mn-atomic layers.

For few-layer MBT with odd-number septuple layers, because the inversion symmetry centered at $P_1$ is preserved, second-order NLO processes are forbidden. While for even-number septuple layers, the cAFM ordering breaks inversion symmetry, making second-order NLOs possible.

However, only breaking inversion symmetry does not guarantee non-zero NLO responses. As shown in Figure 1(c) for bi-septuple layers of MBT, the first-principles calculated electronic bands are spin degenerated and exhibit the $k$ to -$k$ odd-parity in the reciprocal space because of the constraint from the preservation of $P_2T$ and $SU(2)$ spin-rotation symmetries. This inversion symmetry of band structures can be further confirmed by the first-principles calculated energy difference between the lowest conduction band (LCB) and the highest valence band (HVB). In Figure 1(d), the energy difference exhibits a six-fold symmetry because of the $SU(2)$ spin-rotation and crystal three-fold rotation symmetries. Since the injection photocurrent is a transport property associated with the odd-parity velocity operator, these symmetric band structures will induce an overall zero injection current despite the broken inversion symmetry of the magnetic space group.

Fortunately, SOC, which is significant and plays a crucial role in topological properties of MBT family, will break $SU(2)$ spin-rotation symmetry while preserve $P_2T$ symmetry. As shown in Figure 1(e), first-principles calculations reveal that the $k$ to -$k$ odd parity in reciprocal space is eliminated after including SOC. This is also evidenced by that the contour plot of energy difference between LCB and HVB is reduced to a three-fold rotation symmetry in Figure 1(f). Importantly, comparing Figures 1(c) and 1(e), we see that strong SOC leads to particularly significant symmetry breakings near the Γ point, which are Van Hove singularities (vHSs) that are expected to strongly impact optical response. Moreover, beyond the band-edge states plotted in Figure 1(e), we find that SOC breaks even-parity of most conduction and valence bands that cover a wide energy range (see the supplementary material [36]). Thus, remarkable second-order NLO responses are expected for a broad energy spectrum.

Among numerous NLO processes, the shift current and injection current are the two dominant second-order photogalvanic mechanisms. Briefly, the shift and injection currents are determined by nondiagonal and diagonal matrix elements of the current operator in Bloch basis, respectively [11,12]. Since the Berry phase and curvature vanish in these $P_2T$ sysmetry MBT materials, the shift-current photogalvanic effect, related to the Berry-phase difference between valence and conduction bands, shall be zero [12,42,47]. Thus, we will focus on the injection current.

*Injection current*: Historically, the terminology of the injection current was proposed earlier than the shift current, namely the quadratic three-band mechanism, to explain the anomalous bulk photovoltaic effect [10–12]. We first present the formula of circularly-polarized injection current, which has been widely observed in non-centrosymmetric materials. The circular component of the injection current is $j^\gamma_{\alpha\beta} = \kappa^\gamma_{\alpha\beta}(\mathbf{E}(\omega) \times \mathbf{E}^*(-\omega))_{\alpha\beta} = \frac{i}{2}(\eta^\gamma_{\alpha\beta} - \eta^\gamma_{\beta\alpha})E_\alpha E_\beta$, in which $\alpha$ and $\beta$ are the polarization directions of light, $\gamma$ is the outgoing current direction, $\kappa^\gamma_{\alpha\beta}$ is the circular element of the injection-current tensor, and $\eta^\gamma_{\alpha\beta}$ is the off-diagonal element of the injection-current tensor. The diagram approach enables the expression of inject current more concise [41]. Figure 2(a) shows the diagram of the injection current up to second-order perturbations. According to the Feynman rule given by Parker *et. al.* [41], we obtain the corresponding circular element of the injection-current tensor with the velocity gauge (see the derivation in the supplementary material [36] )

$$\kappa^\gamma_{\alpha\beta} = \frac{i\pi e^3}{\hbar^2 \omega_{in}^2} \sum_{vc} \int d^3k \left(v^\alpha_{vc}(k)v^\beta_{cv}(k) - v^\beta_{vc}(k)v^\alpha_{cv}(k)\right)\left(\frac{v^\gamma_{vv}(k)}{\omega - \omega_{vv} - i\xi_{vv}} - \frac{v^\gamma_{cc}(k)}{\omega - \omega_{cc} - i\xi_{cc}}\right)\delta(\omega_{in} - \omega_{cv}) \quad (1)$$

where $v^\alpha_{cv}$ and $v^\gamma_{vv}$ ($v^\gamma_{cc}$) are the interband and intraband velocity matrix elements, respectively, $\omega_{vv} = \omega_v - \omega_v$ and $\omega_{cc} = \omega_c - \omega_c$ are required by the zero-frequency current frequency in Figure 2(a), and $\xi_{vv} = \frac{1}{\tau_v}$ and $\xi_{cc} = \frac{1}{\tau_c}$ are the imaginary part of quasiparticle self-energy, namely the inverse of the lifetime ($\tau$) of the quasi-electron and quasi-hole, respectively. If employing the relaxation-time approximation with the particle-hole symmetry i.e. $\tau_c = \tau_v$, the above expression of injection current is the same as those derived from the density matrix method [11,36] or the polarization-operator method under the length gauge [42]. If further using the two-band model [17] and summation over eigenstates for Berry curvature [48,49], the circular component tensor in Eq. (1) can be written as

$$\kappa^\gamma_{\alpha\beta} = \frac{\pi e^3}{\hbar^2}i\int d^3k\, \Omega^\mu(k)\left(\frac{v^\gamma_{vv}(k)}{\omega - \omega_{vv} - i\xi_{vv}} - \frac{v^\gamma_{cc}(k)}{\omega - \omega_{cc} - i\xi_{cc}}\right)\delta(\omega_{in} - \omega_{cv}), \quad (2)$$

where $\Omega^\mu(k)$ is Berry curvature, and μ is the direction normal to the $\alpha\beta$ plane.

Different from the circularly-polarized case, the injection current under linearly-polarized light is $j^\gamma_{\alpha\beta} = \eta^\gamma(E_\alpha\cos\theta + E_\beta\sin\theta)^2 = \eta^\gamma_{\alpha\alpha}E_\alpha^2\cos^2\theta + \eta^\gamma_{\beta\beta}E_\beta^2\sin^2\theta + 2\eta^\gamma_{\beta\alpha}E_\alpha E_\beta\cos\theta\sin\theta$. Using the relaxation-time approximation with the particle-hole symmetry, diagonal and off-diagonal elements of the current tensor under linearly polarized light is (see derivation details in the supplementary material [36] )

$$\eta_{\alpha\beta}^{\gamma} = \frac{\pi e^3}{\hbar^2 \omega_{in}^2} \sum_{vc} \int d^3k \, (v_{vc}^{\alpha}(k)v_{cv}^{\beta}(k) + v_{vc}^{\beta}(k)v_{cv}^{\alpha}(k))\left(v_{cc}^{\gamma}(k) - v_{vv}^{\gamma}(k)\right) \tau \, \delta(\omega_{in} - \omega_{cv})$$

$$\eta_{\alpha\alpha}^{\gamma} = \frac{2\pi e^3}{\hbar^2 \omega_{in}^2} \sum_{vc} \int d^3k \, v_{vc}^{\alpha}(k)v_{cv}^{\alpha}(k)\left(v_{cc}^{\gamma}(k) - v_{vv}^{\gamma}(k)\right) \tau \, \delta(\omega_{in} - \omega_{cv}) \quad (3)$$

For widely studied polar materials, which hold time-reversal symmetry while breaking spatial inversion symmetry, both Berry curvature $\Omega^{\mu}(k)$ and velocity matrix element $v_{mm}^{\gamma}(k)$ are odd, i.e., $\Omega^{\mu}(k) = -\Omega^{\mu}(-k)$ and $v_{mm}^{\gamma}(k) = -v_{mm}^{\gamma}(-k)$ [11,35]. Therefore, their circular component (Eq. 2) is non-zero while their linear component (Eq. 3) is always zero because of the term, $\frac{1}{2}(v_{vc}^{\alpha}(k)v_{cv}^{\beta}(k) + v_{vc}^{\beta}(k)v_{cv}^{\alpha}(k))$, is real and even. This conclusion agrees with the results obtained by the Glass model [10,50], quantum density operator method [11], and the polarization-operator method [42]. Thus, the injection current is usually called the circular photogalvanic effect in time-invariant non-centrosymmetric systems.

However, the above discussion is not suitable for the materials owning the $PT$ symmetry and strong SOC, e.g., MBT materials. Figures 2(b) and 2(c) show the first-principles calculated $x$-direction intraband velocity matrix with and without SOC for bi-septuple layer MBT, respectively. It is clear to see the antisymmetry of the intraband velocity matrix element without SOC in Figure 2(b). The reason is that the $P_2T$ and $SU(2)$ spin-rotation symmetries enforce intraband velocity matrix elements to be real and antisymmetric (odd parity) in reciprocal space for spin-degenerated bands, i.e. $v_{mm}^{\gamma}(k) = -v_{mm}^{\gamma}(-k)$. While after considering SOC, the $SU(2)$ spin-rotation symmetry is broken. As shown in Figures 2(c) and 2(d), the odd parity of $x$ and $y$ velocity elements is no longer guaranteed, resulting in a non-zero injection current under linearly polarized light.

Interestingly, the circularly-polarized injection current is zero in even-number septuple layers of MBT. The photocurrent illuminated by the circularly polarized light is $j^{\gamma} = \eta_{xx}^{\gamma}E_x^2 + \eta_{yy}^{\gamma}E_y^2 \pm i\kappa_{xy}^{\gamma}E_xE_y$, where $\pm$ denotes left hand (+) or right hand (-) circular polarized light. Regarding $\eta_{xx}^{\gamma} = -\eta_{yy}^{\gamma}$ and $\kappa_{xy}^{\gamma} = 0$ enforced by the 3-fold rotational and $P_2T$ symmetries, photocurrent excited by left-hand or right-hand circularly polarized light should be zero for any direction-current measurement.

Table 1 summarizes the injection current tensors under different polarizations, magnetic orders, and septuple layer numbers of MBTs. Among these cases, the only non-zero photogalvanic effect is the linearly-polarized injection current in even-layer AFM MBTs, which are the zQAH systems that exhibit axion insulator states.

After the symmetry analysis, we employ first-principles calculations to obtain quantitative injection-current spectra. Here we only consider the in-plane polarization because of the known depolarization effect [51]. Figure 3(a) shows that the calculated spectra of a few characteristic components of the in-plane injection-current tensor $\eta_{\alpha\beta}^{\gamma}$ for bi-septuple layers of MBT under linearly polarized light. $\eta_{xx}^{X}$ and $\eta_{yy}^{X}$ are nonzero and significant for a wide range of energy regime starting from the bandgap around 0.14 eV. Moreover, the combination of three-fold rotational and $P_2T$ symmetries leads to $\eta_{xx}^{X} = -\eta_{xy}^{Y} = -\eta_{yy}^{X}$. On the other hand, because of the intraband velocity $v_{nn}^{Y}$ is antisymmetric according to the $\Gamma K$ line in Figure 2(d), the rest inject-current components $\eta_{xx}^{Y} = -\eta_{xy}^{X} = -\eta_{yy}^{Y} = 0$. For example, the zero $\eta_{xx}^{Y}$ is indicated by the black dash-line in Figure 3(a).

It has to be pointed out that the calculation of the above inject-current spectra needs estimated carrier lifetimes (Eq. 3). Unfortunately, there is no quantitative measurement of MBT materials. We notice that 2D semiconductors and thin-film TIs, such as $MoS_2$ [52,53] and $Bi_2Se_3$, typically exhibit carrier lifetimes around 1 ps at room temperature [54]. Particularly, because the band-edge states of MBT are from $p$ orbitals of Bi and Te or Se atoms, we expect that its transport behaviors are similar to those of $Bi_2Se_3$. Therefore, we choose a safe and conservative value, $\tau = 0.1$ ps.

With the parameter, the linearly-polarized photocurrent conductivity can reach $1000 \frac{\mu A}{V^2}$ within the visible-light frequency range, as shown in Figure 3(a) for the family of bi-septuple MBT materials, including $MnBi_2Te_4$, $MnBi_2Se_4$, $MnSb_2Te_4$, $NiBi_2Te_4$, and $NiBi_2Se_4$. Compared with those of notable bulk photovoltaic ferroelectrics, such as $LiNbO_3$ [8,9], $BiFeO_3$ [55–57], and SbSI [58], these values of bi-septuple MBTs are about two orders of magnitude larger, as summarized in Figure 3(b) (see the spectra of other MBT family materials in Supplementary Material [36]). Actually, more aggressive choices of carrier lifetime can further enlarge estimated values of injection current.

Finally, we must address that the linearly-polarized photocurrent in those time-reversal ferroelectrics materials has a different physical origin. Their inject current is zero under linear-polarized light, and photocurrent is from the shift-current mechanism, which is usually much weaker than the injection-current mechanism [12,13,57,59]. In this sense, the time-inversion symmetry broken MBT family has the unique advantage of using linearly-polarized light for NLO applications.

Such giant linearly polarized photocurrent conductivity is due to large SOC in zQAH materials. Because the injection current conductivity is the sum of all allowed transitions with weight of velocity (Eq. 3), both broken symmetries of band-structure and velocity matrix elements in reciprocal space can contribute to non-vanish photocurrent. Figures 3(c) and 3(d) show the first-principles calculated photoconductivity $\eta_{xx}^{X}$ distribution in $k$-space without and with SOC, respectively. We see the exactly antisymmetric (even) photoconductivity in reciprocal space if we do not consider SOC (Figure 3(c) is an example for photon energy $\hbar\omega = 1$ eV). However, in Figure 3(d), the photoconductivity symmetry in reciprocal space is profoundly broken after including SOC (for the photon energy $\hbar\omega = 0.75$ eV, where the photoconductivity reaches to the highest value in Figure 3(a)).

*SHG response*: The features of SHG are similar to the injection-current photogalvanic effect because of similar symmetry requirements. In the following, we focus on two-band and three-band contributions and do not consider the 'Drude weight dipole' diagram [41] that does not involve electron excitations, namely electron transitions between different bands (see details in the Supplemental material [36]). The expression for the SHG susceptibility under the velocity gauge by using the Feynman rule is [36,41]

$$\chi_{\alpha\beta}^{\mu}(2\omega_{in}; \omega_{in}, \omega_{in})$$

$$= \frac{ie^3}{\hbar^2} \sum_{mn} \int \frac{d^3k}{(2\pi)^3} f_{mn} \frac{\{v_{mn}^{\alpha} \ v_{nm;\mu}^{\beta}\}}{(\omega_{in} - \omega_{mn})\omega_{mn}^3} + f_{mn} \frac{4 v_{mn;\beta}^{\alpha} v_{nm}^{\mu}}{(2\omega_{in} - \omega_{mn})\omega_{mn}^3}$$

$$+ \frac{ie^3}{\hbar^2} \sum_{mnl} \int \frac{d^3k}{(2\pi)^3} \frac{\{v_{mn}^{\alpha} \ v_{nl}^{\beta}\}v_{lm}^{\mu}}{2(\omega_{mn} - \omega_{nl})\omega_{lm}\ \omega_{nm}\omega_{ln}} \left( \frac{2f_{ml}}{2\omega_{in} - \omega_{lm}} + \frac{f_{nl}}{\omega_{in} - \omega_{ln}} \right.$$

$$\left. - \frac{f_{mn}}{\omega_{in} - \omega_{nm}} \right), \qquad (4)$$

where $\alpha$ and $\beta$ are the polarization directions of the incident light, and $\mu$ is the polarization direction of emission light. $\{v_{mn}^{\alpha} \ v_{nm;\mu}^{\beta}\}$ is defined as $\frac{1}{2}\left(v_{mn}^{\alpha}v_{nm;\mu}^{\beta} + v_{mn}^{\beta}v_{nm;\mu}^{\alpha}\right)$. The two-photon vertex contributes to the "generalized derivatives" of the velocity matrix element $v_{nm;\mu}^{\alpha}$ [41,42,60], where $v_{nm;\mu}^{\beta} \equiv \frac{\partial v_{nm}^{\beta}}{\partial k^{\mu}} - i[\xi_{nn}^{\mu} - \xi_{mm}^{\mu}]v_{nm}^{\beta}$. Here $\xi_{nn}^{\mu}$ is the Berry connection of the band $n$. Although this generalized derivatives is gauge-independent, it is inconvenient to numerically get the phases of the Bloch functions through the Brillouin zone. In our calculation, we adopt the sum rule $v_{nm;\mu}^{\beta} = -\frac{v_{nm}^{\beta}(v_{nn}^{\mu}-v_{mm}^{\mu})}{\omega_{nm}} - \sum_{l}\left(\frac{v_{nl}^{\beta}v_{lm}^{\mu}}{\omega_{nl}} - \frac{v_{nl}^{\mu}v_{lm}^{\beta}}{\omega_{lm}}\right)$ with a large number of unoccupied bands (300 conduction bands) for converged results [36].

Figures 4(a) and 4(b) show the first-principles calculated real and imaginary parts of SHG susceptibility tensor of bi-septuple MBT, respectively. Because of the three-fold rotational and $P_2T$ symmetries, there are only two independent tensor elements, $\chi_{xx}^{X} = -\chi_{yy}^{X} = -\chi_{xy}^{Y} = -\chi_{yx}^{Y}$ and $\chi_{xx}^{Y} = \chi_{xy}^{X} = \chi_{yx}^{X} = -\chi_{yy}^{Y}$. Moreover, the $\chi_{xx}^{Y}$ is zero because the y-direction velocity matrix is antisymmetric according to the mirror plane ($\Gamma K$ line) shown in Figure 2(d). As a result, only one nontrivial tensor element survives.

The bi-septuple AFM MBT exhibits enhanced SHG. For example, the absolute magnitude of the SHG susceptibility ($\chi_{xx}^{X}$) can reach $28.2 \times 10^{-6}$ $esu$ at 0.13 eV, as shown in Figure 4 (c). This is nearly one order magnitude higher than that of monolayer h-BN [61,62], MoS$_2$ [61–66], and bilayer AFM CrI$_3$ [21]. To reveal the role of SOC in SHG responses, we calculated the SHG susceptibility tensor by tuning the strength of SOC. In Figure 4 (c), the magnitude of the SHG susceptibility tensor is decreased by the strength of SOC and finally disappeared when SOC is zero. This result agrees with the theoretical analysis, i.e., SOC breaks the antisymmetry of velocity matrix elements and the symmetry of band structure, enabling SHG.

SHG responses can be characterized by shedding a linearly polarized laser beam onto materials and measuring the response of different polarization components of the incident light. By inspecting its angular dependence, *e.g.*, rotating the sample or the direction of the polarization of incident light, the crystallographic orientation and SHG polarization anisotropy and intensity can be determined. Figures 4(d) and 4(e) show the components of outgoing SHG response for different polarized directions at the photon frequency ω = 0.13 eV. The *Y*-direction outgoing SHG response has a 45-degree rotation with respect to that along the *X*-direction. This is due to $\chi^X_{xx} = -\chi^Y_{xy}$. Interestingly, the outgoing SHG response is zero for the circularly polarized incident light. This vanishing circularly-polarized SHG is from the conservation of the three-fold rotational and $P_2T$ symmetries, similar to the reason for the zero circular injection current. It is worth mentioning that the SHG response at high frequencies is much smaller than that at low frequencies, which is different from the injection current. This is because the SHG response described in Eq. 4 has an extra factor $\frac{1}{\omega_{in}}$, resulting in a faster decay for higher frequencies.

To summarize, we discover the giant photogalvanic effect and SHG under linearly polarized light for the even-number septuple layers MBT family materials, i.e., the zQAH systems. The origin of this giant second-order photo-response is the *PT*-symmetry and large SOC in topological materials. The disappeared photoresponse under circularly polarized light is because of the combination of *PT*-symmetry and three-fold rotation symmetry. This unique polarization condition and enhanced NLOs shed light on the novel detector and optoelectronic applications based on topological magnetic materials.

**Acknowledgment**

R. Fei thanks Dr. Linyuan Gao for helpful discussions. This work is supported by the National Science Foundation (NSF) CAREER grant No. DMR-1455346, NSF EFRI2DARE-1542815, and the Air Force Office of Scientific Research (AFOSR) grant No. FA9550-17-1-0304. The computational resources are provided by the Stampede of Teragrid at the Texas Advanced Computing Center (TACC) through XSEDE.

**Figures:**

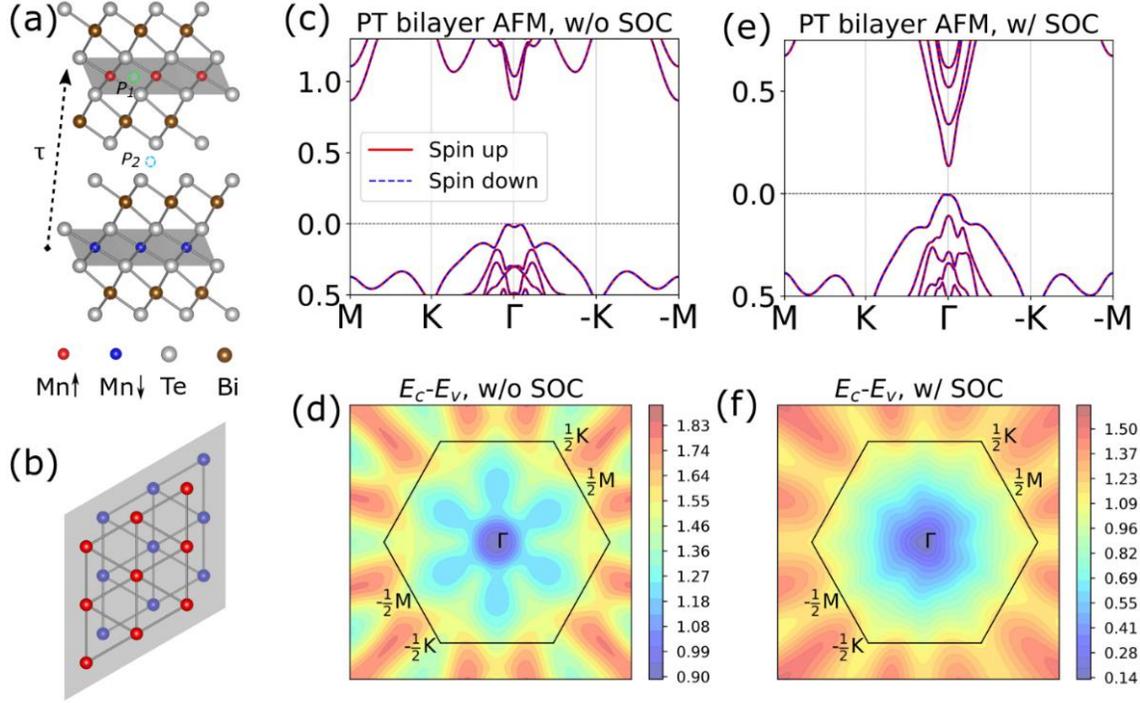

**Figure 1** (a) and (b) Side and top views of the atomic structure of AFM MBT crystal. $\tau$ is the half translation, $P_1$ is the inversion center of odd-number-layer AFM MBT, and $P_2$ is the inversion center for even-number-layer MBT. (c) and (d) The band structure without and with SOC, respectively. (d) and (e) The energy difference between the LCB and HVB without and with SOC, respectively.

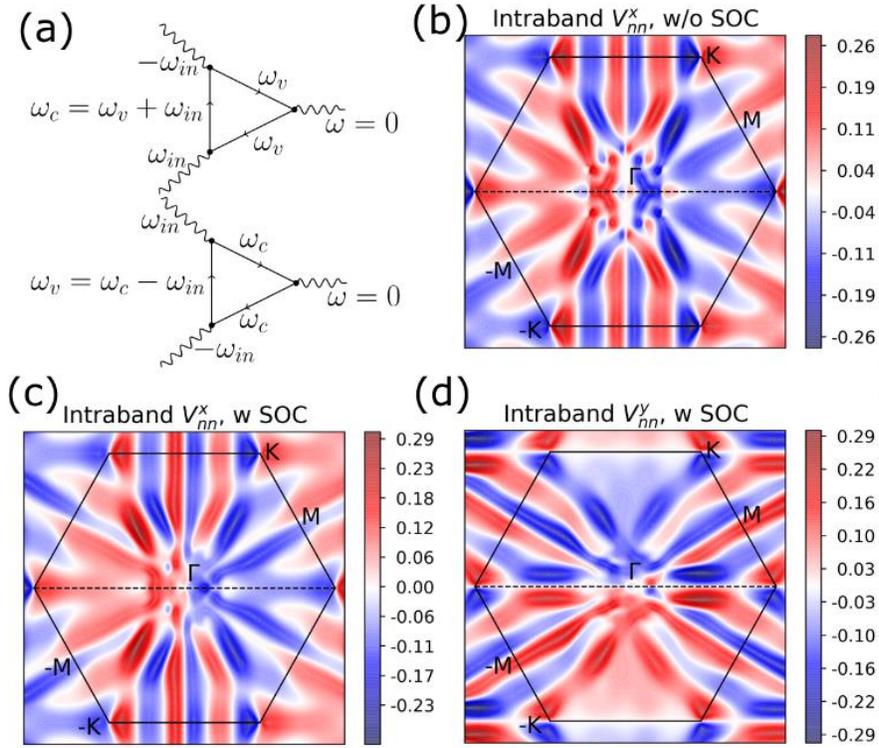

**Figure 2** (a) The Feynman diagram for injection current. (b) and (c) The first-principles calculated intraband *x*-direction velocity matrix element without and with SOC, respectively. (d) That along the *y* direction with SOC.

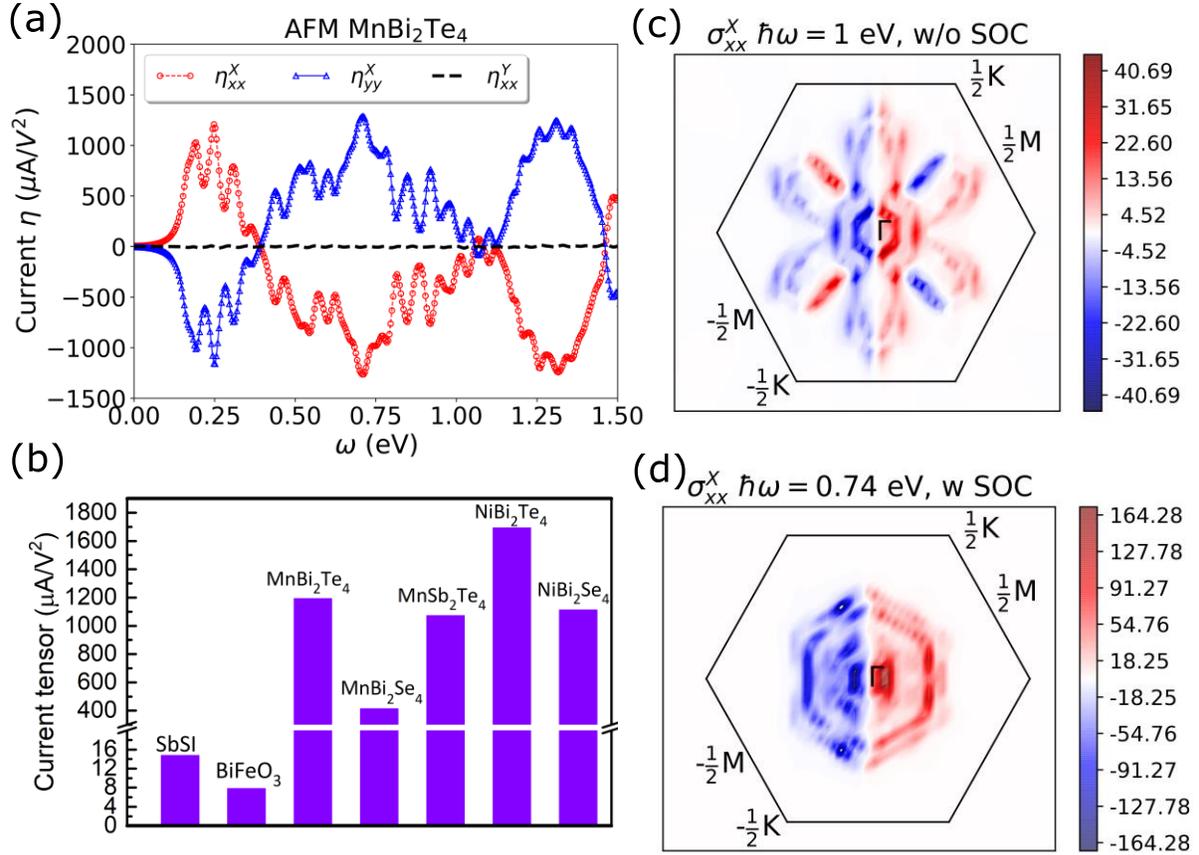

**Figure 3** (a) The linearly-polarized injection-current matrix element spectrum for bilayer MBT. (b) The maximal photocurrent response under linearly-polarized light for different MBT materials, bulk SbSI, [58] and BiFeO3 [55–57]. (c) and (d) The contour plots of first-principles calculated photoconductivity $\eta_{xx}^{X}$ in the receiprical space without and with SOC, respectively.

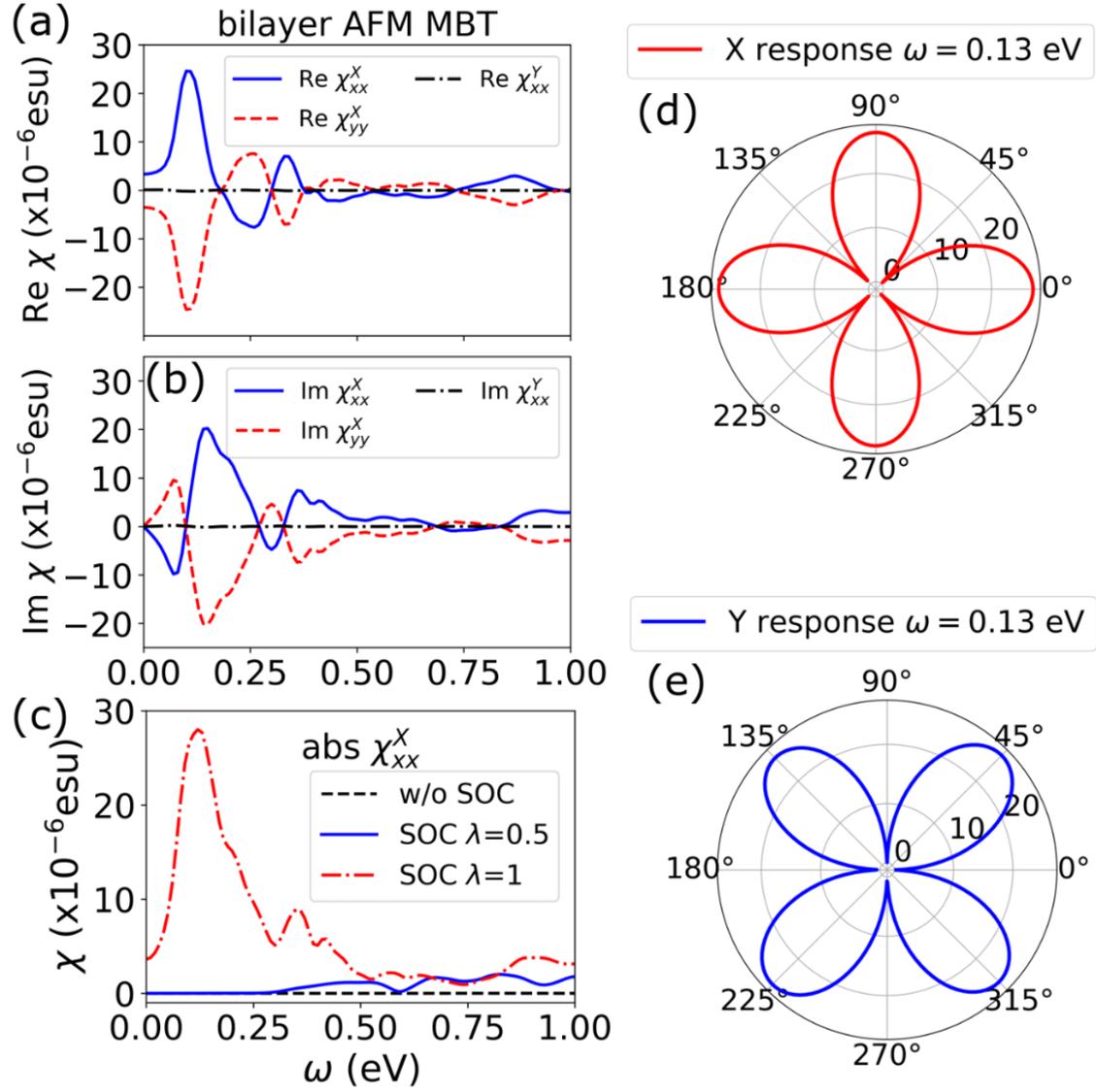

**Figure 4** (a) and (b) The real and imaginary parts of the SHG susceptibility tensor, respectively. (c) The absolute SHG susceptibility tensor $\chi_{xx}^X$ for different SOC strength λ. λ = 1 indicates the realistic case. (d) and (e) The $x$-direction and y-direction polarization component of the outgoing SHG response for different polarized incident photons at ω = 0.13 eV.

**Table 1** The interband velocity matrixes $v_{mn}^{\alpha}(k)v_{nm}^{\alpha}(k)$, intraband velocity matrix $v_{mm}^{\gamma}$, Berry curvature $\Omega^z(k)$, the linearly-polarized inject-current matrix element $\eta^{\gamma}$, circular inject-current matrix element $\kappa^{\gamma}$ for different magnetic states (AFM and FM), and the parity of the layer number (even or odd). The symmetry reasons for the parity of some matrixes are listed inside the square brackets.

|  | Layers | Interband velocity matrixes $v_{mn}^{\alpha}(k)v_{nm}^{\alpha}(k)$ | Intraband velocity matrix $v_{mm}^{\gamma}(k)$ | Berry curvature $\Omega^z(k)$ | Linearly-polarized inject current $\eta_{\alpha\alpha}^{\gamma}$ | Circular inject-current $\kappa_{\alpha\beta}^{\gamma}$ ($\alpha \neq \beta$) |
|---|---|---|---|---|---|---|
| AFM | even | not symmetric | not asymmetric [SOC] | zero [$P_2 T$] | non-zero | zero |
|  | odd | symmetric | asymmetric [$P_1$] | symmetric [$P_1$] | zero | zero |
| FM | even | symmetric | asymmetric [$P_2$] | symmetric [$P_2$] | zero | zero |
|  | odd | symmetric | asymmetric [$P_1$] | symmetric [$P_1$] | zero | zero |